\documentclass{emulateapj}
\usepackage{amsmath}
\usepackage{color}

\newcommand\ba{\begin{eqnarray}}
\newcommand\ea{\end{eqnarray}}

\usepackage{hyperref}
\hypersetup{
colorlinks=true,
linkcolor=red,
citecolor=blue,
filecolor=cyan,
urlcolor=magenta,
}

\usepackage{graphicx}
\usepackage{natbib}
\usepackage{multirow}
\usepackage{subfigure}

\begin{document}
 
\title[Constraint on Light Dipole Dark Matter from Helioseismology]{
Constraint on Light Dipole Dark Matter from Helioseismology}
\author{Il\'\i dio Lopes~\altaffilmark{1,2}, Kenji Kadota~\altaffilmark{3}, 
and Joseph Silk~\altaffilmark{4,5,6}}

\altaffiltext{1}{Centro Multidisciplinar de Astrof\'{\i}sica, Instituto Superior T\'ecnico, 
Universidade de Lisboa , Av. Rovisco Pais, 1049-001 Lisboa, Portugal;ilidio.lopes@ist.utl.pt,ilopes@uevora.pt} 
\altaffiltext{2}{Departamento de F\'\i sica,Escola de Ciencia e Tecnologia, 
Universidade de \'Evora, Col\'egio Luis Ant\'onio Verney, 7002-554 \'Evora, Portugal} 
\altaffiltext{3}{Department of Physics, Nagoya University, Nagoya 464-8602, Japan;kadota.kenji@f.nagoya-u.jp}
\altaffiltext{4}{Institut d'Astrophysique de Paris, UMR 7095 CNRS, Universit\'e Pierre et Marie Curie, 98 bis Boulevard Arago,  F-75014 Paris, France; silk@astro.ox.ac.uk} 
\altaffiltext{5}{Department of Physics and Astronomy, The Johns Hopkins University, Baltimore, MD21218, USA} 
\altaffiltext{6}{Department of Physics, University of Oxford, UK} 

\date{\today}

\begin{abstract} 
We investigate the effects of a magnetic dipole
moment of asymmetric dark matter (DM) in the evolution of the Sun.
The dipole interaction can lead to a sizable DM scattering cross section even for light DM, 
and asymmetric DM can lead to a large DM number density in the Sun.
We find that solar model precision tests,  using
as diagnostic  the sound speed profile obtained from helioseismology data,
exclude dipolar DM particles with a mass  larger than  $4.3\;{\rm GeV} $ 
and magnetic dipole moment larger than  $1.6\times10^{-17} \;{\rm e\; cm } $. 
\end{abstract}


\maketitle
%
%
\section{Introduction}

The universe  
is composed of baryons and  unknown nonbaryonic particles, 
commonly called dark matter. 
Although the gravitational interaction between baryons and DM is well established,
other types of interactions with the standard particles 
are less well known.    

Several experiments seek to detect the DM particle  
by observing its scattering off nuclei, by detecting some by-product resulting 
from its annihilation into high energy particles, or by producing them  
in accelerators through the collision of standard particles.
The goal is for one or more of these experiments to obtain a signal 
of the DM interaction, other than the well-known gravitational interaction.
The first positive hints of direct DM observations are intriguing, 
although  controversial:      
DAMA/LIBRA~\citep{2010EPJC...67...39B}, 
CoGeNT~\citep{2011PhRvL.106m1301A}, CRESST~\citep{2012EPJC...72.1971A} 
and CDMS~\citep{2013arXiv1304.4279C} collaborations report indications of positive signals 
which do not comply with standard explanations of weak interacting massive particle  interactions.
Furthermore, other collaborations, such as
XENON~\citep{2012PhRvL.109r1301A} and CDMS~\citep{2013PhRvD..88c1104A}, can nearly exclude 
the positive results found by the previous experiments.

This experimental data has stimulated interest in light DM $\lesssim 10$ GeV as a candidate.
For the purpose of illustrating the potentially stringent constraints on  light DM, we consider an operator which can induce large enough DM-nuclear interactions even for small momentum transfer due to a small DM mass. One of the simplest extensions of the standard model for this purpose would be the dipole operator which is the only gauge invariant operator up to dimension five, letting fermionic DM with an intrinsic dipole moment couple to the photons ~\citep{1994PhLB..320...99B,2006PhRvD..73h9903S,2009PhRvD..80c6009M,2010PhLB..693..255H}. 
Such magnetic dipole dark matter (MDDM) can successfully explain  recent claims of
direct detections, including DAMA/LIBRA and CoGeNT~\citep{2010PhRvD..82b3533A,2011PhLB..696...74B}. 
Moreover, several constraints have been set on MDDM
using  direct search data~\citep{2011PhRvD..83h3510L,2012JCAP...08..010D},
astrophysical and cosmological observations~\citep{2013arXiv1307.7120F} and the 
Large Hadron Collider data~\citep{2012PhRvD..85f3506F,2012PhLB..717..219B}. 
  
Here, we use helioseismic data to set constraints on
asymmetric MDDM.
In particular, we  focus on the solar sound speed radial profile,
for which there is a reliable inversion obtained from seismic 
data~\citep{2009ApJ...699.1403B,1997SoPh..175..247T}.
  
\begin{figure}
\centering
\includegraphics[scale=0.5]{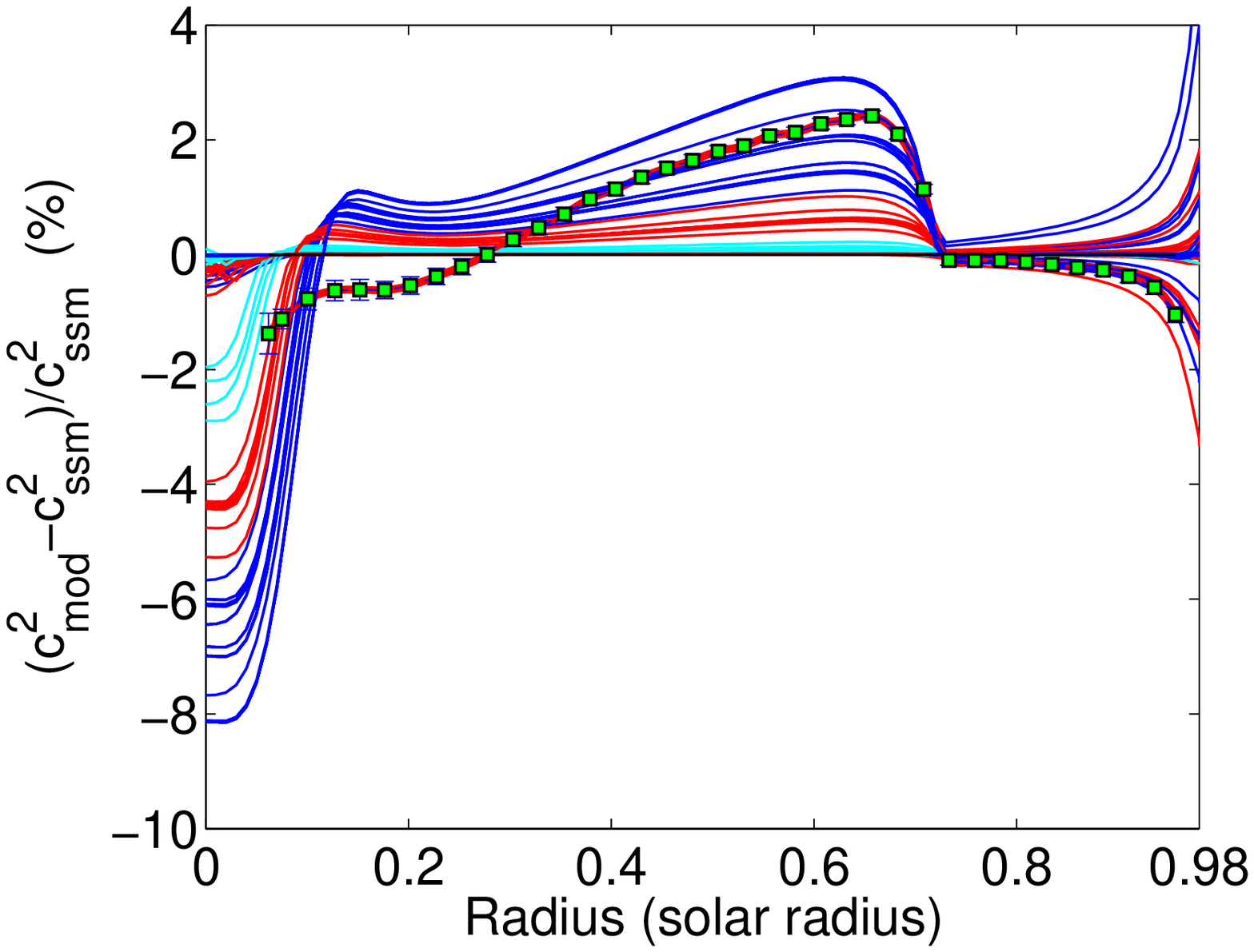}
\caption{
\label{fig1}
Comparison of the sound speed radial profile between the SSM~\citep{2013ApJ...765...14L} and different solar models evolving within  an environment rich in MDDM.
The red-dotted-green curve corresponds to the difference between inverted 
sound speed profile~\citep{1997SoPh..175..247T,2009ApJ...699.1403B} 
and our SSM~\citep{1993ApJ...408..347T,2013ApJ...765...14L}.
The continuous curves correspond to  DM particles  that have a mass $m_\chi$ of  
$1$ -- $20\; {\rm GeV }$ 
(blue curve $m_\chi \le 8\; {\rm GeV }$, red curve $8 \le m_\chi \le 12\; {\rm GeV }$ and cyan curve $m_\chi \ge 12\; {\rm GeV }$)
and a magnetic dipole that takes values from  $10^{-15}\;{\rm e\; cm }$ to  $10^{-19}\;{\rm e\; cm }$. 
In the core of the Sun, the variation caused by the presence of MDDM 
 is much larger that the current sound speed 
difference  between theory and observation.}
\end{figure}

\section{Properties of Magnetic Dipole Dark matter}

This study focuses  on the DM fermion $\chi$ which couples to the photon by the magnetic dipole interaction Lagrangian ${\cal L}_{\rm MDDM}=\frac{\mu_{\chi}}{2} \bar{\chi} \sigma_{\mu \nu} F^{\mu \nu} \chi$ where $ F^{\mu \nu}$ is the electromagnetic field 
strength tensor, $ \sigma_{\mu \nu}$ is the commutator of two Dirac matrices
and  $\mu_{\chi}$ is the magnetic dipole moment. We notice this interaction 
vanishes for Majorana fermions, so the fermionic DM particle has to be Dirac. 

The fermionic DM particle with a magnetic dipole moment arises
in many models of DM, including among others models related to technicolor~\citep[e.g.,][]{2007PhRvD..76e5005F}. 
In particular, we are interested in the case of the interaction of DM with the baryons 
that takes place by means of a massless mediator to yield   long-range interaction, 
a process quite distinct from the contact-like interaction~\citep{2012IJMPA..2750065D,2012JCAP...08..010D}. 
The photon is the  obvious mediator, among more exotic candidates such as the dark photon~\citep{2011PhRvD..84k5002F,2011JHEP...02..100C}.    

The DM differential scattering cross section with respect to the nuclei recoil energy $E_R$ off the nucleus with a spin $I$
with a DM particle of spin $S_\chi$, via the electromagnetic interaction 
between the nuclear magnetic moment $\mu_{Z,A}$
(of a nucleus of mass $A$ and charge $Z$) and the magnetic moment $\mu_\chi$ 
of the DM particle, is described by~\citep{2011PhLB..696...74B}:
\begin{eqnarray}
\frac{d\sigma_{\chi}}{dE_R}
=\frac{e^2\mu_\chi^2}{4\pi E_R}\frac{S_\chi+1}{3S_\chi} 
\left[ {\cal A}_{E} |G_E|^2 + {\cal B}_{M} |G_M|^2   \right]
\label{eq:1}
\end{eqnarray}
where  ${\cal A}_{E}$ and ${\cal B}_{M}$ are the electrical and magnetic moment terms,
and $ G_E(E_R)$ and $ G_M(E_R)$ are the nuclear form factors (normalized
to  $G_E(0)=1,G_M(0)=1$) to take into account the
elastic scattering off a heavy nucleus.
${\cal A}_{E}$  and ${\cal B}_{M}$ read
\begin{eqnarray}
{\cal A}_{E} =
Z^2\left(1-\frac{E_R}{2m_A v_r^2} - \frac{E_R}{m_\chi v_r^2} \right) 
\label{eq:2}
\\
{\cal B}_{M} =
\frac{I+1}{3I}\left(\frac{\mu_{Z,A}}{e/(2 m_p)} \right)^2 \frac{m_A E_R}{m_p^2 v_r^2},
\label{eq:3}
\end{eqnarray}   
where $e$ and $m_p$ are the elementary electric charge and the mass of the proton~\citep{2012PhLB..717..219B}. 
$m_A$ and $m_\chi$ are the masses of the baryon nucleus and DM
particle, respectively. $E_R$ and $v_r$ are defined from the transfer momentum ${\bf q}$ and
the center-of-mass momentum ${\bf p}$, such that ${\bf q}^2=2m_A E_R$ and $|{\bf p}|=m_r v_r$
where $m_r=m_A m_\chi/(m_A + m_\chi)$.

At first glance, it is reasonable to expect that the contribution of heavy elements  for the interaction inside the
Sun could be quite significant, due to the strong dependence in $Z$ and $I$ (see Equations~(\ref{eq:1})--(\ref{eq:3})). However, because metals inside the
Sun contribute to less than 2\% of the total mass of the Sun, their contribution can be considered to be negligible.
The electromagnetic interaction between MDDM and baryons will be important 
for hydrogen and helium   which correspond to 98\% of the total mass of the Sun,
from which Helium abundance in the Sun's core changes during its evolution 
from an initial abundance of 27\% to 62\% for the current age of the Sun. 
Nevertheless, this increase of helium occurs only in the very center of the Sun, 
the atomic number of helium ($Z=2$) is relatively small  (when compared with other chemical elements), 
and the helium abundance is always smaller than that of hydrogen.
Furthermore, during the Sun's evolution, to a  good approximation, it is reasonable to consider only the interaction  of MDDM with hydrogen to be relevant.
The total elastic scattering cross section of DM
with hydrogen $\sigma_{\chi p}$, from
Equations~(\ref{eq:1})--(\ref{eq:3}) with a uniform form factor, can then be expressed as 
$\sigma_{\chi p}^{\rm SI} + \sigma_{\chi p}^{\rm  SD}$ 
where $\sigma_{\chi p}^{\rm SD}$ and $\sigma_{\chi p}^{\rm  SI}$ are the conventional effective scattering cross sections ~\citep{2012PhLB..717..219B}:
\begin{eqnarray}
\sigma_{\chi p}^{\rm  SI} =\frac{\mu_{\chi}^2\;e^2}{2\pi}  
\left(1-\frac{m_r^2}{2m_p^2}-\frac{m_r^2}{m_pm_\chi}\right)
\label{eq:sigmapSI}
\end{eqnarray}
and
\begin{eqnarray}
\sigma_{\chi p}^{\rm SD} =\frac{\mu_{\chi}^2\;e^2}{2\pi} 
\left(\frac{\mu_p}{e/ (2 m_p)}\right)^2\;\frac{m_r^2}{m_p^2}, 
\label{eq:sigmapSD}
\end{eqnarray}
with $\mu_p$ being the magnetic moment of the
proton. Both $\sigma_{\chi p}^{\rm  SD}$ and $\sigma_{\chi p}^{\rm  SI}$ contribute equally for the total scattering cross section, $\sigma_{\chi p}$.
However, in the case that the mass of the DM particle is much larger than the
mass of the proton,i.e., $m_\chi \gg m_p$, it follows that   $m_r\approx m_p$
and the  $\sigma_{\chi p}$ becomes independent of the mass of the DM particle,
and $\sigma_{\chi p}$ is only  proportional to the square of the DM magnetic moment,
$\sigma_{\chi p}\approx 8.3 \; e^2/(2\pi)\;\mu_\chi^2$.   
 
\begin{figure}
\centering
\includegraphics[scale=0.5]{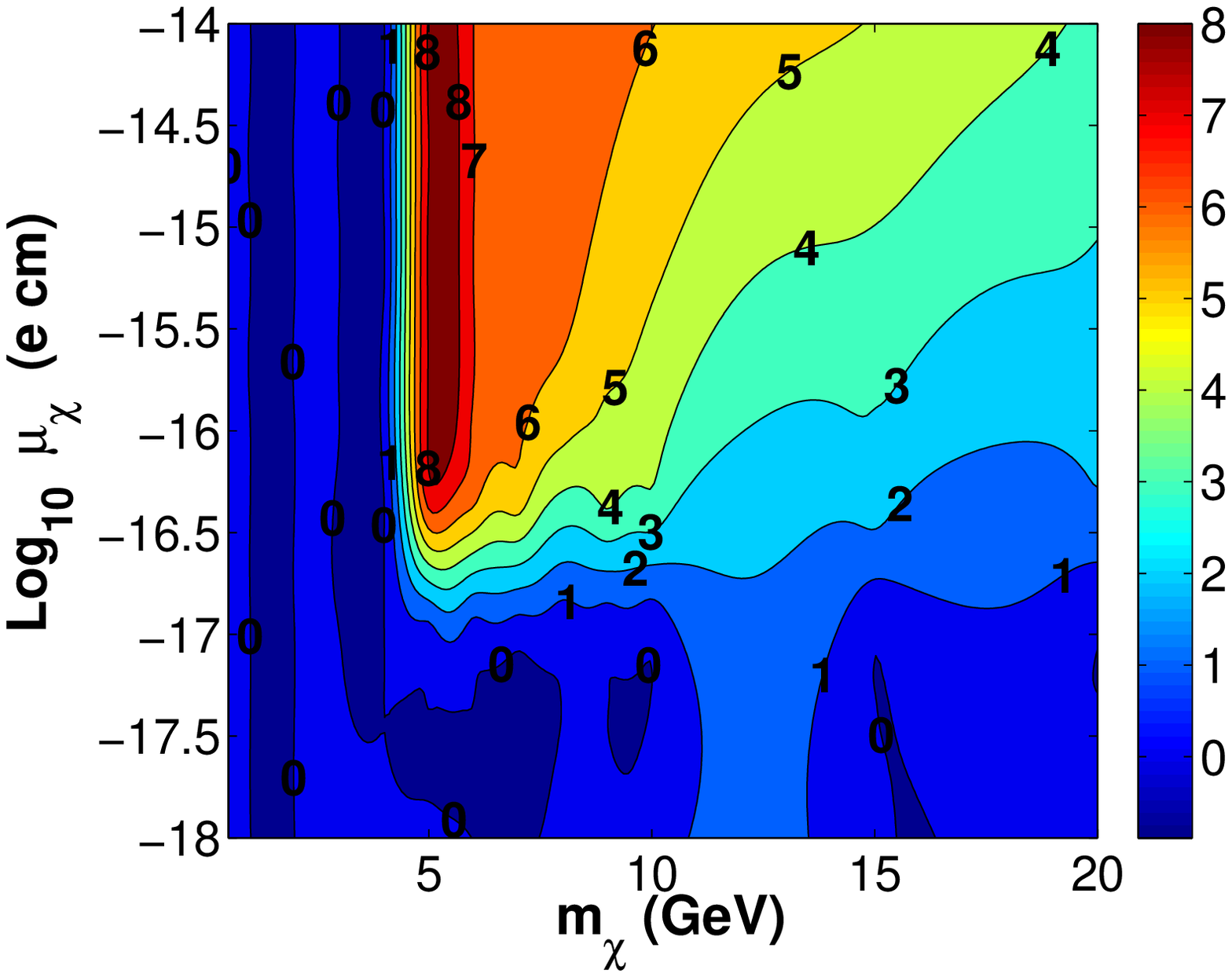}
\caption{
\label{fig2}
Exclusion plot for magnetic dipole DM parameter space ($m_\chi$ -- $\mu_\chi$)
from present day low-Z SSM and helioseismology data.
The possible candidates must lie in the light region, above the iso-contour
with $2\%$.  The different isocontour curves represent the maximum difference, i.e., 
$\max{\left[(c_{\rm mod}^2-c_{\rm ssm}^2)/c_{\rm ssm}^2\right]}$ 
in the region below $0.3\;R_\odot$ 
-- the percentage of the 
maximum sound difference between the SSM and the MDDM solar models.
The MDDM halo is assumed to be an isothermal sphere 
with local density $\rho_\chi = 0.38\; {\rm GeVcm^{-3}}$, and  thermal velocity (dispersion)  
$v_{\rm th} = 270 \; {\rm km s^{-1}}$.}
\end{figure}

\section{Dark matter and the standard solar model}
%
Our Galaxy and the Sun are assumed to be immersed in an isothermal sphere 
composed of MDDM particles. The observational determination
of the DM density $\rho_{\chi}$ in the neighborhood of the Sun is quite uncertain,  
varying from $0.3$ to $0.85 \; {\rm GeV cm^{-3}}$~\citep{2012ApJ...756...89B,2012MNRAS.tmp.3493G}. 
Here we set $\rho_\chi =  0.38 \; {\rm GeV cm^{-3}} $~\citep{2010JCAP...08..004C}. 
 
DM particles flow through any celestial object such as the Sun.
A few of these particles will occasionally scatter with the
atomic nuclei of the star, mostly with protons, losing energy
in the process, in some cases the energy reduction is such that
the velocity of the DM particle becomes smaller than the Sun's
escape velocity, i.e., the DM particle becomes gravitationally
bound to the star.
As time passes, the DM particle undergoes more collisions, 
until the particle reaches thermal equilibrium with local baryons. 
 
The accretion of DM by the Sun  depends on the balance between
three processes: {\it capture} of DM via  energy loss of the colliding
particle with the baryon nuclei; {\it annihilation} of two DM particles into 
standard particles; and {\it evaporation} of DM particles, important 
for light DM particles for which the collisions with nuclei result in
escape from the Sun's gravitational field. Therefore, the total number of
DM particles inside the Sun is  determined by 
the relative magnitude of these three processes.  

At each step of the Sun's evolution,  
the total number of particles $N_\chi$  
that accumulate inside the star is computed  by solving 
the following equation numerically~\citep{2012ApJ...757..130L}:
\begin{eqnarray}
\frac{dN_{\chi}(t)}{dt}=C_{c}-C_{a} N_{\chi}(t)^2 -C_{e}N_{\chi}(t),
\label{eq-Nchi}
\end{eqnarray}
where $C_{c}$ is the rate of capture of particles from the MDDM halo,  
$C_{a}$ is the annihilation rate 
of particles, and $C_{e}$ in the evaporation rate of DM particles 
from the star.
The capture rate $C_{c}$  is computed numerically
at each step of the evolution  from the expression obtained by~\citet{1987ApJ...321..571G}. 
The cross section used in the computation of 
$C_{c}$ corresponds to the ones given by Equations~(\ref{eq:sigmapSI})--(\ref{eq:sigmapSD}). 
In the following, we will consider the asymmetric DM scenario and hence $C_{a}$ can be neglected. A concrete realization could  include asymmetric composite DM which can induce a sizable dipole for the bound state of the particles \citep{2007PhRvD..76e5005F,2008PhRvD..78k5010R,2011PhRvD..84b7301D}.

The evaporation of DM particles is not  relevant for $m_\chi \ge 4\;{\rm GeV }$,
nevertheless for less massive particles, the effect of evaporation is
not negligible and the capture process goes into equilibrium with the evaporation.
In the regime where the Sun is optically  
thin with respect to  DM particles,~\citet{2013JCAP...07..010B} we
have obtained an approximate expression for $C_e$,  which we  use in our calculations 
to define the lower bound of less massive DM particles. 
This expression reproduces the full numerical results  
with an accuracy better than 15\%~\citep{2011NuPhB.850..505K}.

The reference solar model used in our computation corresponds to the standard solar model (SSM),  
which has been updated with the most recent physics~\citep{1993ApJ...408..347T}. 
In particular, we choose to use the solar composition determined by~\citet{2009ARA&A..47..481A},
usually known as the low-metallicity (or low-Z) SSM, as discussed by~\citet{2013ARA&A..51...21H}.
Figure~\ref{fig1} shows the sound speed difference of reference 
that corresponds to the sound speed difference between the low-Z SSM
and the sound speed obtained by inversion of  the seismic data of the
Global Oscillations at Low Frequencies (GOLF) and Michelson Doppler Image (MDI)  
instruments of the Solar and Heliospheric Observatory 
Satellite and BiSON observational network~\citep{1997SoPh..175..247T,2009ApJ...699.1403B}.
A detailed discussion about the physics 
of our SSM can be found in~\citet{2013ApJ...765...14L}.
This SSM is identical to others low-Z SSM published in the 
literature~\citep[e.g.,][]{2013PhRvD..87d3001S}. 

The solar models evolving in different MDDM halos are obtained by a similar procedure to the
SSM. Likewise these models are required to have the observed solar 
radius and luminosity at the present age.   
In our description of the impact of   DM 
on the evolution of the Sun, we closely follow  recent developments in this  
field~\citep{2010PhRvD..82j3503C,2011PhRvD..83f3521L,2012ApJ...752..129L,2012ApJ...757..130L,2013ApJ...765L..21C}. A detailed description 
of how this process is implemented  in our code is discussed in~\citet{2011PhRvD..83f3521L}. 

The accumulation of MDDM particles inside the Sun
reduces the  temperature in the Sun's core   and as a consequence, the sound speed drops, 
but is compensated for by an increase of sound speed in the
radiative region and the convection zone (see Figure~\ref{fig1}). 
This results from  the fact  that these solar models are required to have 
a radius and luminosity consistent with observations.
The calibration follows an iterative procedure identical to the one used 
to compute the SSM. 
In principle, we could use the sound speed and density profiles 
obtained from inversion of helioseismology data as a diagnostic tool, however, 
we prefer to use the sound speed because only frequencies of acoustic modes are observed, 
consequently the sound speed inversion is the more reliable diagnostic method. 
In the future, if frequencies of gravity modes are measured with success, 
the density profile can become an independent method to probe the Sun’s core.
Figure~\ref{fig1} shows that the sound speed differences of the solar models
computed for different values of  $m_\chi$ and $\mu_\chi$
are quite distinct from the sound speed difference of reference. 
This effect is more important for DM particles of relatively 
low mass and high magnetic moment. In the case of particles with a very low 
$m_\chi$ the impact on the sound speed difference profile becomes 
insignificant due to the occurrence of  DM evaporation. 
Although  the DM affects  the whole internal structure of the star equally,
we focus our analysis on the Sun's core where the direct impact of DM is detected.
It is reasonable to consider that for solar models for which the sound speed difference is larger 
than the sound speed difference of the reference model, or equivalently if this difference  
is larger than 2\%, then these solar models can be excluded on the basis that  they cannot be accommodated with our current understanding of the physics of the solar interior.  
It is true that in the Sun's deep core, the sound speed difference of the reference solar model 
still contains  a few uncertainties     coming either  from an insufficient description 
of the physics of   the SSM, or poor inversion of the sound speed profile 
due to lack of low degree seismic data.
It is believed that some of the current problems in the SSM is related 
with abundances and opacities below the base of the convection zone, 
but these localized uncertainties do not affect the core of the Sun 
where this diagnostic is done. Moreover, their effect on the 
Sun's structure will be smaller than the observational sound speed difference. 
Nevertheless, this uncertainty is at most of the order of 1.5\%. 
Alternatively, if we choose to use as reference a high-Z SSM, 
as the sound speed difference with observations is of the order of 0.3\%, 
the constraint on the MDDM parameters could be stronger. Nevertheless, 
due  to the problem related with the chemical composition 
in the solar interior~\citep{2011ApJ...743...24S}, 
we take the conservative approach to use as reference  
the low-Z SSM which has the largest observational uncertainty.

Figure~\ref{fig2} shows the MDDM exclusion plot computed for different values
of $m_\chi$ and  $\mu_\chi$. We choose as diagnostic the value corresponding to the
maximum difference between the square of the sound speed of the SSM
and the sound speed of the DM solar models. There is a region of the parameter
space for the relatively light DM $4.0 \le m_\chi\le 20.0\; {\rm GeV} $ and 
with magnetic moment $  \mu_\chi \ge 10^{-17}\;{\rm e cm} $ for which 
the sound speed difference is larger than $2\%$. Accordingly, these models
can be rejected.  
We found the quantitatively same exclusion limits on the MDDM parameters even if we use the density profile, rather than the sound speed profile, as a diagnostic method for which 
the observational density uncertainty is considered to be of the order of 4\%.

\section{Summary and Conclusion}

In this Letter, we use helioseismology to constrain the mass 
and magnetic moment of MDDM.  We find that there
is an important set of parameters for which the impact of
MDDM is much larger than the current difference
between theory and observation.   
We have found that solar model precision tests using the sound speed profile obtained from helioseismology data
as a diagnostic exclude MDDM with  $m_\chi \ge 4.3\;{\rm GeV} $ and 
$\mu_\chi \ge 1.6\times10^{-17} \;{\rm e\; cm } $.
DM particles with the above parameters produce changes in the Sun 
which are much larger than the current sound speed difference between theory and observation. 

Furthermore, this new constraint does not affect the results found by~\citet{2012JCAP...08..010D}
for which a DM particle with $m_\chi\sim 10\;{\rm GeV }$ 
and $\mu_\chi\sim 10^{-18}\;{\rm e\; cm }$ could accommodate the positive signal detections 
of  DAMA, CoGeNT and CRESST experiments and simultaneously not be ruled out by
CDMS, XENON and PICASSO. Our solar constraint result slightly improves the bound on the dipolar DM from the Large Hadron Collider (LHC) and Large Electron–Positron Collider (LEP): 
LHC using the mono-jet events plus missing transverse energy puts the bound $\mu_{\chi} \lesssim 10^{-15}\;{\rm e\; cm }$ 
for $m_{\chi}=10$ GeV and LEP using the mono-photon plus missing transverse energy events improves the 
bound to $\mu_{\chi} \lesssim 10^{-16}\;{\rm e\; cm }$~\citep{2012PhRvD..85f3506F}.

For the Sun, unlike in the case of other stars, 
the stellar parameters are very well-known. Nonetheless, 
there is an uncertainty in parameter determination related to
the nature of the DM particles and their interaction with baryons, 
namely $m_\chi$ and $v_{\rm th}$. Nevertheless, this uncertainty should not 
produce   variation in the capture rate larger than  15\%~\citep{2011PhRvD..83f3521L}. 
A major concern comes from the DM density used in the calculation,
for which one of the more recent measurement predicts a value 2.2 times larger 
than the value used~\citep{2012ApJ...756...89B,2012MNRAS.tmp.3493G}. Accordingly,  
the capture rate of DM particles by the Sun increases by an  order of magnitude.
Therefore, it is expected that the iso-contours could change slightly. 
   
\begin{acknowledgments}
The authors thank the anonymous referees for their positive comments and questions.
IL and KK thank the hospitality of IAP where this work was initiated.
The research of IL  has been supported by grants from "Funda\c c\~ao para a Ci\^encia e Tecnologia" 
and "Funda\c c\~ao Calouste Gulbenkian". The research of KK has been supported by Grant-in-Aid for Scientific Research from the MEXT of Japan.  The research of JS has been supported at IAP by ERC project  267117 (DARK) hosted by Universit\'e Pierre et Marie Curie - Paris 6 and at JHU by NSF grant OIA-1124403.
\end{acknowledgments}
%


\end{document}